\documentstyle{article}

\makeatletter

\twocolumn
\def\@normalsize{\@setsize\normalsize{12pt}\xpt\@xpt
\abovedisplayskip 10pt plus2pt minus5pt\belowdisplayskip \abovedisplayskip
\abovedisplayshortskip \z@ plus3pt\belowdisplayshortskip 6pt plus3pt
minus3pt\let\@listi\@listI} 

\def\subsize{\@setsize\subsize{12pt}\xipt\@xipt}

\def\section{\@startsection {section}{1}{\z@}{24pt plus 2pt minus 2pt}
{12pt plus 2pt minus 2pt}{\large\bf}}

\def\subsection{\@startsection {subsection}{2}{\z@}{12pt plus 2pt minus 2pt}
{12pt plus 2pt minus 2pt}{\subsize\bf}}

\def\maketitle{
 \date{}
 \par
 \begingroup
 \def\thefootnote{\fnsymbol{footnote}}
 \def\@makefnmark{\hbox
 to 0pt{$^{\@thefnmark}$\hss}}
 \if@twocolumn
 \twocolumn[\@maketitle]
 \else \newpage
 \global\@topnum\z@ \@maketitle \fi\thispagestyle{plain}\@thanks
 \endgroup
 \setcounter{footnote}{0}
 \let\maketitle\relax
 \let\@maketitle\relax
 \gdef\@thanks{}\gdef\@author{}\gdef\@title{}\let\thanks\relax
 \thispagestyle{empty}
 \relax}
\def\@maketitle{\newpage
 \null
 \vskip .65in \begin{center}
 {\LARGE \@title \par} \vskip .35in {\large \lineskip .5em
\begin{tabular}[t]{c}\@author
 \end{tabular}\par}
 \vskip 1em {\large \@date} \end{center}
 \par
 \vskip .25in}

\makeatother

\newcommand{\COMMENT}[1]{}

\newtheorem{theorem}{Theorem}
\newtheorem{proposition}[theorem]{Proposition}

\newtheorem{corollary}[theorem]{Corollary}
\newtheorem{definitions}{Definitions}
\newtheorem{definition}[definitions]{Definition}

\newcommand{\qed}{\vspace{.1em}\noindent\fbox{\rule{
0em}{.1em}\rule{.1em}{0em}}\vspace{1em}}

\newcommand{\Min}{\mbox{\sc Min}}
\newcommand{\Max}{\mbox{\sc Max}}

\newcommand{\game}{M}
\newcommand{\rows}{r}
\newcommand{\cols}{c}
\newcommand{\row}{i}
\newcommand{\col}{j}
\newcommand{\gameMin}{\game_{\min}}
\newcommand{\gameMax}{\game_{\max}}
\newcommand{\minDist}{p}
\newcommand{\maxDist}{q}
\newcommand{\minUniform}[1]{{\cal P}_{#1}}

\newcommand{\compMeas}{\game}
\newcommand{\compMeasMin}{\compMeas_{\min}}
\newcommand{\compMeasMax}{\compMeas_{\max}}

\newcommand{\progClass}{{\cal P}}
\newcommand{\inpClass}{{\cal I}}

\newcommand{\circComp}{{\cal C}}
\newcommand{\progComp}{{\cal P}}

\newcommand{\gameOracle}{\game}

\newcommand{\val}[1]{{\cal V}({#1})}

\newcommand{\poly}{P}
\newcommand{\sigmaTwo}{\Sigma_2^P}
\newcommand{\piTwo}{\Pi_2^P}
\newcommand{\sigmaTwoOracle}{\Sigma_2^{\poly(\gameOracle)}}

\newenvironment{proof}{
\noindent{\bf Proof}\ }{
\qed

}

\begin{document}

\title{\huge Simple Strategies for Large Zero-Sum Games
  \\ \huge with Applications to Complexity Theory}

\author{
  Richard J.~Lipton~\thanks{
    Computer Science Dept, Princeton Univ., Princeton, NJ 08544.
    Supported in part by NSF grant CCR-9304718.
    Email: rjl@cs.princeton.edu}
  \\ Princeton University
  \and
  Neal E.~Young~\thanks{
    Dept.~of Operations Research and Ind.~Eng.,
    Cornell University, Ithaca, NY 14853.
    This work was partly supported by
    NSF grants CCR-8906949 and CCR-9111348
    and \'Eva Tardos' PYI grant,
    and partly done at 
    UMIACS, University of Maryland, College Park, MD 20742.
    Email: ney@orie.cornell.edu.
    \protect\vspace{1in}
    }
  \\ Cornell University
}

\maketitle

\begin{abstract}
\small

Von Neumann's Min-Max Theorem 
guarantees that each player of a zero-sum matrix game
has an optimal mixed strategy.
We show that
each player has a {\em near-optimal} mixed strategy
that chooses {\em uniformly} from a multiset 
of pure strategies of size logarithmic in the number
of pure strategies available to the opponent.
Thus, for exponentially large games,
for which even {\em representing} an optimal mixed strategy 
can require exponential space,
there are near-optimal, linear-size strategies.
These strategies are easy to play
and serve as small witnesses to the approximate value of the game.

Because of the fundamental role of games,
we expect this theorem to have many applications in complexity theory
and cryptography.
We use it to strengthen the connection established by Yao
between randomized and distributional complexity 
and to obtain the following results:
(1) Every language has {\em anti-checkers} --- small hard multisets of
inputs certifying that small circuits can't decide the language.
(2) Circuits of a given size can generate random instances 
that are hard for all circuits of linearly smaller size.
(3) Given an oracle $\gameOracle$ for any exponentially large game,
the approximate value of the game
and near-optimal strategies for it can be computed in $\sigmaTwoOracle$.
(4) For any NP-complete language $L$,
the problems of (a) computing a hard distribution of instances of $L$
and (b) estimating the circuit complexity of $L$ 
are both in $\sigmaTwo$.
\end{abstract}

\section{Introduction}
Games play a fundamental role in many parts of theory. For example,
cryptographic problems can often be viewed as games between those who wish to
keep a secret and those who wish to discover it
\cite{Goldwasser:Micali:Rackoff:89}. Many computational classes can be defined
in natural ways as games: for example, PSPACE can be defined
in this way \cite{Chandra:Kozen:Stockmeyer:81}.
Other times games arise in a slightly more subtle way. For example, questions
about how hard it is to generate hard instances of some problem can be modeled
as a game between the generator and the algorithm. Yao \cite{Yao:77,Yao:83}
exploits this idea to prove lower bounds on randomized algorithms.

The classic result on games is the famous Min-Max Theorem of von Neumann
\cite{vonNeumann:28}, which guarantees that each player of a zero-sum game has
an optimal mixed strategy.  For exponentially large games, optimal strategies
are generally exponentially large.  In many cases, we need to know not only
that an object exists but also that it is not too complex. Without this latter
restriction we cannot use the object.

\paragraph{Simple strategies for large games.}
Our first result is a variant of von Neumann's Min-Max Theorem that shows
that each player has a near-optimal mixed strategy that plays uniformly
from a multiset of size logarithmic in the
number of pure strategies available to the opponent.  
The proof is a surprisingly simple probabilistic argument
similar to circuit derandomization techniques \cite{Adleman:78,Schoning:87}.  
However, the central nature of games in theory
suggests that this simple result may have far-reaching consequences.  
This result was obtained independently by Alth\"ofer \cite{Althofer:94}.

\paragraph{Strengthening the connection 
between randomized and distributional complexities.}
This connection was first established by Yao \cite{Yao:77,Yao:83}.
He considered a game where $\Min$'s pure strategies 
are the deterministic algorithms in a given class,
$\Max$'s pure strategies are the inputs of a given size,
and the payoff for a particular pair 
is the cost of the algorithm on the input.
If $\Min$ moves first, the expected payoff
can be interpreted as the worst-case expected complexity
of the best randomized algorithm.
If $\Max$ moves first, the expected payoff
can be interpreted as the average-case complexity
of the best algorithm for the hardest input distribution.
By von Neumann's theorem, these are the same.
Thus, the worst-case complexity of the best randomized algorithm
equals the optimal average-case complexity against the
hardest input distribution.

A main drawback is that for equality to hold, 
the ``randomized algorithms'' must generally
be allowed to have exponentially large encodings.  
Because of this, Yao's theorem has been used mainly in the weaker direction:
to prove lower bounds on randomized complexity
and upper bounds on average-case complexity.
The stronger direction (equality) holds only 
for complexity measures that allow program size
to grow {\em exponentially} with input size.

Our variant of the Min-Max theorem reduces the dependence on encoding size.
Our variant implies that it suffices to consider
randomized algorithms that have {\em linear-size} encodings.
Thus, the stronger direction holds (approximately)
for complexity measures that allow program size
to grow linearly with input size.
This includes most measures of circuit complexity.

Note that this application is similar to known
circuit derandomization techniques \cite{Adleman:78,Schoning:87}.
However, the theorem has many other applications.
For instance, by applying it to the program/input game
for the {\em input} player we show that there are hard distributions
that can be {\em generated} by small circuits.

\paragraph{Anti-checkers and circuits that generate hard random instances.}
We give applications concerning the
complexity of generating and solving hard random instances of problems.
Our main application is to show that every language has {\em anti-checkers}
--- small multisets of inputs
such that correctly classifying a fraction of the inputs
in the multiset is nearly as hard as correctly classifying
{\em all} inputs of the given size.  
Circuits of a given size can use anti-checkers to generate
random instances that are hard for all slightly smaller circuits.

\paragraph{Uniform complexity.}
We obtain related results for uniform complexity measures.
Specifically, we show that
the following problems are in $\sigmaTwo$:
\begin{itemize}
\item estimating the value of any exponentially large game
  given an oracle for the payoffs;
\item computing approximate upper and lower bounds 
  on the circuit complexity of $L$ and
\item computing hard random instances of $L$,
\end{itemize}
where $L$ is any NP-complete language.

\section{Other Related Work}

Theorem  \ref{simple min-max}, 
our first variant of von Neumann's Min-Max Theorem,
was obtained independently by Alth\"ofer \cite{Althofer:94}.
He considers applications to other linear programs,
large game trees, and uniform sampling spaces.

A subsequent work \cite{Young:94}
gives simple greedy algorithms 
that (given the payoff matrix)
find the $k$-uniform strategies
shown to exist in Theorems \ref{simple min-max} 
and \ref{dovetailed min-max}.

\paragraph{Uniform complexity.}
As mentioned previously, the complexity class PSPACE
has a natural characterization via games.
More recently, the complexity classes NEXP and coNEXP
have been similarly characterized
\cite{Feigenbaum:Koller:Shor:93}.
Our variant of von Neumann's Theorem can be used in
these characterizations of NEXP and coNEXP.

Most research on hard distributions to date concerns uniform complexity.
A significant body of work concerns average-case {\em completeness}, e.g, 
\cite{Gurevich:87,Venkatesan:Levin:88,Gurevich:90,%
Impagliazzo:Levin:90,Ben-David:Chor:Goldreich:Luby:92}.
These results are analogous to NP-completeness results,
except they concern distributional problems
--- $\langle$problem, input distribution$\rangle$ pairs.
These results relate the complexities of classes of distributional problems.
Generally, few relations to worst-case complexity are known
(see, however, \cite{Ben-David:Chor:Goldreich:Luby:92}).

Ben-David et al.~\cite{Ben-David:Chor:Goldreich:Luby:92}
and Li and Vitanyi \cite{Li:Vitanyi:89} 
show the existence of distributions under which 
the average-case complexity of any program 
is within a constant (exponential in the size of the program)
of the worst-case complexity.
Generating random instances from such distributions is difficult
--- it requires diagonalizing against all programs in question.
The result applies to uniform complexity classes, not circuits.
More precisely, it generates inputs that are hard only for
programs that are exponentially smaller than the inputs.

\paragraph{Circuit complexity.}
Schapire \cite{Schapire:89} shows that his technique
for boosting the correctness of PAC-learning strategies 
can also be applied to boost the correctness of circuits.
This implies the existence of distributions from which random
instances are nearly as hard for circuits as worst-case instances.
His results establish a version of Corollary \ref{error cor},
weaker in that the complexity of {\em generating} the distribution
is not known and in that the factor in bound (\ref{error cor 1}) 
is a larger polynomial.

\paragraph{Upper bounds.}
One example of the use of the Min-Max Theorem in the stronger direction
(to upper bound randomized complexity,
measured, in this case, by the competitive ratio)
is given by Alon, Karp, Peleg and West \cite{Alon:Karp:Peleg:West:92}.  
They show the existence of randomized $k$-server strategies 
by considering a certain zero-sum matrix game.  
The competitiveness of the strategy
is related to the value of the game, which in turn depends
on the underlying metric space.

\section{Simple Strategies}
\label{min-max sec}
A {\em two-player zero-sum game} 
is specified by an $\rows\times \cols$ matrix $\game$ 
and is played as follows.
$\Min$, the row player, 
chooses a probability distribution $\minDist$ over the rows.
$\Max$, the column player, 
chooses a probability distribution $\maxDist$ over the columns.
A row $\row$ and a column $\col$ 
are drawn randomly from $\minDist$ and $\maxDist$,
and $\Min$ pays $\game_{\row\col}$ to $\Max$.
$\Min$ plays to minimize the expected payment; $\Max$ plays to maximize it.
The rows and columns are called the {\em pure strategies}
available to $\Min$ and $\Max$, respectively,
while the possible choices of $\minDist$ and $\maxDist$ 
are called {\em mixed strategies}.
The Min-Max Theorem states that playing first
and revealing one's mixed strategy is not a disadvantage:
\begin{theorem}[\cite{vonNeumann:28}]
\label{min-max}
\[
\min_\minDist \max_\col \sum_{\row} \minDist(\row) \game_{\row\col}
=
\max_\maxDist \min_\row \sum_{\col} \maxDist(\col) \game_{\row\col}
\]
\end{theorem}
Note that the second player need not play a mixed strategy
--- once the first player's strategy is fixed, 
the expected payoff is optimized for the second player
by some pure strategy.
The expected payoff when both players play optimally
is called the {\em value} of the game.  We denote it $\val{\game}$.

\subsection{Simple strategies for large games.}
Games that model computations are often exponentially large.
Generally, the optimal strategies are the primal and dual solutions, 
respectively, to an $O(NM)$-size linear program.
For exponentially large games, 
optimal strategies are generally too large to even represent.
This motivates considering smaller mixed strategies:
\begin{definition}
A mixed strategy is {\em $k$-uniform}
if it chooses uniformly from a multiset of $k$ pure strategies.
\end{definition}

We show that for $k$ proportional to the {\em logarithm} 
of the number of pure strategies available to the opponent,
each player has a near-optimal $k$-uniform strategy.

Let $\gameMin$ and $\gameMax$ 
denote $\min_{\row\col} \game_{\row\col}$ 
and $\max_{\row\col} \game_{\row\col}$,
respectively.
Recall that $\game$ is an $\rows\times\cols$ matrix.

\begin{theorem} \label{simple min-max}
  For any $\epsilon>0$ and $k \ge \ln (\cols)\,/\,2\epsilon^2$,
  \begin{displaymath}
    \min_{\minDist\in \minUniform{k}} 
    \max_\col \sum_{\row} \minDist(\row) \game_{\row\col}
    \le \val{\game} + \epsilon(\gameMax - \gameMin),
  \end{displaymath}
  where $\minUniform{k}$ denotes the $k$-uniform strategies for $\Min$.
  Equality holds only if $k = \ln (\cols)\,/\,2\epsilon^2$.
  The symmetric result holds for $\Max$.
\end{theorem}
\begin{proof}
  Assume WLOG that $\gameMin = 0$ and $\gameMax = 1$.
  Fix $\epsilon>0$ and $k > \ln (\cols)\,/\,2\epsilon^2$,
  and form $S$ by drawing $k$ times independently at random
  from $\Min$'s optimal mixed strategy.
  For any fixed pure strategy $\col$ of the opponent,
  the probability that 
  \begin{equation}  \label{smm bound}
    \sum_{\row\in S} \frac{1}{|S|}\game_{\row\col} \ge \val{\game}+\epsilon
  \end{equation}
  is bounded by $e^{-2k\epsilon^2}$.
  This is because the left-hand side is the average 
  of $k$ independent random variables in $[0,1]$
  with expected value at most $\val{\game}$ \cite{Hoeffding:63}.

  By the choice of $k$, $e^{-2k\epsilon^2} < 1/\cols$.
  Thus, the expected number of the opponent's $\cols$ pure strategies
  that satisfy (\ref{smm bound}) is less than $1$.
  Since the number of such strategies is an integer,
  it must be zero for {\em some} $S$ of size $k$.
\end{proof}

For many important games, $\gameMax-\gameMin$ is constant.
In this case, the theorem says that for any $\epsilon$,
$\Min$ has an $O(\log \cols)$-uniform strategy
that is within $\epsilon$ of optimal.

To model {\em dovetailing} computations,
we give the following variant, 
in which $\Min$ plays a small subset of pure strategies 
(called a {\em dovetailing set}) simultaneously,
choosing the best once $\Max$ commits to a play.

\begin{theorem}  \label{dovetailed min-max}
  For $\epsilon>0$ and $k \ge \log_{1+\epsilon} \cols$,
  \begin{displaymath}
    \min_{|S|=k} \max_\col \min_{\row\in S} \game_{\row\col}
    \le \val{\game} + \epsilon (\val{\game} - \gameMin).
  \end{displaymath}
  Equality holds only if $k = \log_{1+\epsilon} \cols$.
  The symmetric result holds for $\Max$.
\end{theorem}
We omit the proof, 
which is similar to the proof of Theorem \ref{dovetailed min-max}.

\section{Distributional vs.~Randomized Complexity}
\label{appl sec}
We next consider Theorems \ref{simple min-max} and \ref{dovetailed min-max}
in the context of the program/input game introduced by Yao.

\begin{definitions}
Fix a finite class $\progClass$ of programs,
a finite class $\inpClass$ of inputs
and a function $\compMeas: \progClass \times \inpClass \rightarrow \Re$
(where $\compMeas(\row,\col)$ represents some cost
of the computation $\row(\col)$).

The {\em (unlimited) randomized complexity} of $\compMeas$ is 
$\min_\minDist \max_{\col\in \inpClass} 
 \sum_\row \minDist(\row) \compMeas(\row,\col)$,
where $\minDist$ ranges over the probability distributions on $\progClass$.

The {\em (unlimited) distributional complexity} of $\compMeas$ is
$\max_\maxDist\min_{\row\in \progClass} 
 \sum_\col \maxDist(\col) \compMeas(\row,\col)$,
where $\maxDist$ ranges over the probability distributions on $\inpClass$.

The {\em program/input} game for $\compMeas$
is the two-player zero-sum game given by
$\game_{\row\col} = \compMeas(\row,\col)$ for $\row\in \progClass$ and $\col\in \inpClass$.
\end{definitions}

As Yao observed, von Neumann's theorem applied to the program/input game 
implies that the unlimited randomized complexity
and the unlimited distributional complexity
are equal to $\val{\compMeas}$.
As a corollary of Theorem \ref{simple min-max} applied for each player,
we obtain the following.

\begin{definitions}
A {\em $k$-uniform randomized program} is a randomized program
obtained by playing uniformly from a multiset of $k$ programs in $\progClass$.

The {\em $k$-uniform randomized complexity} of $\compMeas$ is
$\min_\minDist\max_{\col\in \inpClass} 
 \sum_\row \minDist(\row) \compMeas(\row,\col)$,
where $\minDist$ ranges over the $k$-uniform distributions on $\progClass$.

The {\em $k$-uniform distributional complexity} of $\compMeas$ is
$\max_\maxDist\min_{\row\in \progClass} 
 \sum_\col \maxDist(\col) \compMeas(\row,\col)$,
where $\maxDist$ ranges over the $k$-uniform distributions on $\inpClass$.

\end{definitions}

\begin{corollary}  \label{simple program/input}
  Let $\Delta = \compMeasMin-\compMeasMax$.
  
  \begin{enumerate}
  \item 
    For any $\epsilon>0$ and $k > \ln (|\inpClass|)\,/\,2\epsilon^2$,
    the $k$-uniform randomized complexity of $\compMeas$
    exceeds the unlimited randomized complexity
    by less than $\epsilon \Delta$.

  \item 
    For any $\epsilon>0$ and $k > \ln (|\progClass|)\,/\,2\epsilon^2$,
    the unlimited distributional complexity of $\compMeas$
    exceeds the $k$-uniform distributional complexity
    by less than $\epsilon \Delta$.
  \end{enumerate}
\end{corollary}

A good $k$-uniform randomized program
corresponds to a multiset of $k$ programs
such that, for any input, 
the average complexity of those programs on that input
is close to the unlimited randomized complexity of $\compMeas$.
A good $k$-uniform input distribution
corresponds to a multiset of $k$ inputs
such that, for any program, 
the average complexity of that program on those inputs
is close to the unlimited distributional complexity.

Sometimes it is also useful to consider 
small sets of programs such that, on any input,
{\em some} program achieves a low complexity on that input.
Similarly, one might want a small set of inputs
such that any program has high complexity 
on {\em at least one} of the inputs in the set.
We call such small sets {\em dovetailing sets}.
As a corollary to Theorem \ref{dovetailed min-max},
we obtain the following.
\begin{corollary}
  \label{dovetailed program/input}

  \begin{enumerate}
    \item 
      For any $\epsilon>0$ and $k > \log_{1+\epsilon} |\inpClass|$,
      there exists a set of at most $k$ programs such that, for any input,
      the complexity of some program in the set is less than
      $\val{\compMeas} + \epsilon(\val{\compMeas} - \compMeasMin)$
      on that input.

    \item 
      For any $\epsilon>0$ and $k > \log_{1+\epsilon} |\progClass|$,
      there exists a set of at most $k$ inputs such that, for any program,
      the complexity of the program
      is more than $\val{\compMeas} - \epsilon(\compMeasMax-\val{\compMeas})$
      on some input in the set.
  \end{enumerate}
\end{corollary}

\section{Anti-checkers against circuits.}
An {\em anti-checker for $L$ against circuits of size $s$}
is a multiset of inputs such that any circuit of size $s$
fails to correctly classify (w.r.t.~$L$)
a fraction of the inputs in the multiset.
Anti-checkers are similar to {\em program checkers} \cite{Blum:Kannan:89}
(which verify program correctness on a {\em per-input} basis)
in that anti-checkers allow certification of the complexity of $L$
on a per-circuit basis.

We apply Corollary \ref{simple program/input} is to show that, 
provided $s$ is slightly less than the circuit size required to decide $L$
without error, there are anti-checkers for $L$ of size $O(s)$.
As a consequence, we obtain small circuits that generate hard random inputs.

Other flavors of anti-checkers for various complexity measures
and with different notions of ``anti-checking'' are possible.
To illustrate the issues,
at the end of this section
we give a variation in which the anti-checker
is a small set of inputs such that any program
of a given size has a high {\em running time}
on {\em at least one} of the inputs in the set.

The first form of anti-checker
is obtained by applying Corollary \ref{simple program/input}
to a program/input game where the programs are the circuits of size $s$
and the inputs are the binary strings of size $n$.
(More generally, we could take the programs to
be those with encoding $\row$ ($0 \le \row < 2^s$)
and the inputs to be those with encoding $\col$  ($0 \le \col < 2^n$).
We require only that the program encoding scheme
satisfy some basic compositional properties.)
We take the complexity measure to be correctness,
i.e., the payoff of the program/input game
is zero if the program is correct on the input and one otherwise.

As described below in the proof of Theorem \ref{anti-checker thm},
a $k$-uniform randomized program
with worst-case probability of error less than $1/2$
yields a deterministic program of size $O(ks)$ that is correct on all inputs.
Thus, for circuits just slightly smaller than the smallest circuit
deciding membership without error,
there are hard input distributions
on which no such circuit achieves a probability of error
significantly less than $1/2$.
Further, there are such hard input distributions
which are $k$-uniform for small $k$.
The underlying multiset yields the desired anti-checker.

\begin{definition}
  Define $\circComp_L$, the {\em circuit complexity} of language $L$, 
  to be the function such that $\circComp_L(n)$ is the size 
  (length of the encoding in binary) of the smallest circuit 
  deciding membership in $L$ of all $n$-bit binary strings.
\end{definition}

\begin{theorem}  \label{anti-checker thm}
  There exists a number $N$ such that,
  for any language $L$ and numbers $n > N$, $\epsilon > 0$,
  and $s \le \circComp_L(n)\epsilon^2\,/\,3n$,
  there exists a multiset of $s/\epsilon^2$ length $n$ binary strings
  such that every circuit of size $s$ misclassifies 
  at least a fraction $1/2-\epsilon$ of the strings in the multiset.
\end{theorem}
\begin{proof}
  Let $\compMeas(\row,\col)$ be 0 if the $\row$th size $s$ circuit
  correctly decides whether the $\col$th $n$-bit binary string is in $L$
  and 1 otherwise.
  Let $\delta = 1/2 - \val{\game}$.
  The two parts of Corollary \ref{simple program/input} respectively imply:
  \begin{enumerate}
  \item[i.]
    There are $1+n\ln (2)\,/\,2\delta^2$ circuits of size $s$
    such that on any $n$-bit string, a majority of the circuits
    classifies the string correctly.

  \item[ii.]
    Provided $\epsilon > \delta$,
    there are $s\ln (2)\,/\,2(\epsilon-\delta)^2$ $n$-bit strings
    such that any size $s$ circuit misclassifies at least
    a fraction $1/2-\epsilon$ of the strings.
  \end{enumerate}
  From (i), it follows that there is a circuit of size
  $ns\ln (2)\,/\,2\delta^2 + s + O(n/\delta^2)$
  that correctly classifies all $n$-bit strings.
  (The circuit returns the majority of what the $n/2\delta^2$ circuits return.)
  Thus, $ns\ln (2)\,/\,2\delta^2 +s+O(n/\delta^2)\ge \circComp_L(n)$.
  By the choice of $s$, this implies
  $\delta/\epsilon \le \sqrt{\ln (2)\,/\,6} + O(1/n)$.
  This implies that, for large enough $n$,
  the number of strings in (ii) is 
  at most $s/\epsilon^2$.
\end{proof}

For instance, taking $\epsilon=1/3$, $n>N$ and $s \le \circComp_L(n)/27n$,
there exists a multiset of $9s$ inputs
such that any circuit of size $s$ 
errs on one sixth of the inputs in the multiset.
Intuitively, the problem of computing all $2^n$ inputs correctly 
is harder than 
the problem of computing a fraction of a fixed multiset of inputs correctly.
Thus, it is surprising that such hard multisets exist.

Note also the contrapositive:
to show that $\circComp_L(n) \le 27ns$,
it suffices to exhibit, for every multiset of $9s$ inputs,
a size $s$ circuit that errs
on less than one sixth of the inputs in the multiset.
Note that the tradeoff here is close to tight:
for any such multiset, 
some circuit of size $O(sn)$ correctly classifies every input in the multiset.

Similar results are possible for other complexity measures
(e.g., running time, space, circuit depth, etc.).
There are three general considerations:
\begin{enumerate}
\item 
  Instead of considering {\em expected} complexity 
  (e.g., the expected running time of a program),
  one considers the {\em probability}  
  that the complexity exceeds a given threshold.
  This yields a game with small $\compMeas_{\max}-\val\compMeas$,
  which allows small anti-checkers.

\item
  For some complexity measures,
  to build a deterministic program that has low complexity on all inputs, 
  it suffices to find a small set of programs such that, on any input,
  {\em at least one} (as opposed to a majority) of the programs in the
  set has low complexity.

\item
  One might be interested in a weaker form of anti-checker,
  one such that any program has high complexity on {\em at least one}
  (as opposed to a fraction) of the inputs in the set.
\end{enumerate}

The following example illustrates these three considerations. 

\begin{definitions}
  Let $\progComp_L(n,t)$ denote the size of the smallest program 
  that decides language $L$ in time $t$ for all $n$-bit inputs.
\end{definitions}

\begin{theorem} \label{dovetailed anti-checker thm}
  Fix any language $L$ and numbers $n$, $t$ and $s < \progComp_L(n,t)$.
  Let $\inpClass$ be the inputs of size $n$;
  let $\progClass$ be the programs of size $s$ 
  that correctly decide $L$ on inputs of size $n$.

  There exists a set $S\subseteq \inpClass$ of size
  \[ O\left(\frac{s}{\log\frac{\progComp_L(n,O(tn))}{ns}}\right) \]
  such that each program in $\progClass$
  requires more than time $t$ on some input in $S$. 
\end{theorem}
\begin{proof}
  For $\row\in \progClass$ and $\col\in \inpClass$,
  let $\compMeas(\row,\col)$ 
  be zero if program $\row$ runs in time $t$ on input $\col$
  and one otherwise.
  The value of the program/input game for $\compMeas$
  is the minimum probability of exceeding time $t$
  by any program on a random input from the hardest input distribution.
  Let this value be $1-\delta$.
  By Corollary \ref{dovetailed program/input}, 
  \begin{enumerate}
    \item[i.]
      Taking $\epsilon = 1/(1-\delta)-1$, for $k = O(n/\delta)$,
      there exists a set of $k$ programs such that, 
      for any input, the complexity of some program in the set
      is less than $1 = (1-\delta)(1+\epsilon)$ on that input.

    \item[ii.]
      Taking $\epsilon=1/\delta - 1$, for $k = O(s/\log(1/\delta))$,
      there exists a size $k$ set of inputs such that, 
      for any program,
      the complexity of the program is more than $0 = 1-\delta(1+\epsilon)$
      on some input in the set.
  \end{enumerate}
  
  By (i), there exists a program of size 
  $O(ns/\delta)$ that correctly classifies each size $n$ input
  in time $O(nt)$.  This program simply dovetails the $k$ programs
  in the set and returns when the first program finishes.
  (At least one of the programs finishes in time $t$.)
  Thus, $\delta = O(sn/\circComp_L(n,O(nt)))$.

  By (ii), there exists a set of $O(s/\log(1/\delta))$ inputs
  such that any program of size $s$ takes time at least $t$
  on at least one of the inputs in the set.
\end{proof}

\subsection{Generating hard random instances.}
An easy corollary of the existence of small anti-checkers
is that circuits of a given size 
(up to the circuit complexity of the language)
can {\em generate} random inputs
that are hard for all slightly smaller circuits
to classify correctly.

\begin{definition}
  For any probability distribution $D$ on $\{0,1\}^*$,
  define $\circComp_{L,D}$,
  {\em the circuit complexity of deciding $\langle L,D \rangle$ with error},
  to be the function such that
  $\circComp_{L,D}(n,\epsilon)$ is the size of the smallest circuit 
  deciding membership in $L$ with probability of error at most $\epsilon$
  when given a random input drawn from $D$
  restricted to strings of size $n$.
\end{definition}

\begin{corollary}
  \label{error cor}
  There exists an $N$ such that,
  for any language $L$, 
  and any numbers $n > N$, $0 < \epsilon \le 1/2$, and $s' \le \circComp_L(n)$,
  some circuit of size $s'$ computes a distribution $D$
  such that
  \COMMENT{\bf reduce $O(s')$ to $O(s'/n)$}
  \begin{eqnarray}
    \label{error cor 1}
    \circComp_{L,D}(n,1/2-\epsilon) & \ge & \Omega(s'\epsilon^2/n) \\
    \label{error cor 2}
    \circComp_{L,D}(n,0) & \le & O(s').
  \end{eqnarray}
\end{corollary}
The proof is a straightforward application of Theorem \ref{anti-checker thm}
--- the circuit computes the uniform distribution
on the anti-checker against circuits of size $\Omega(s'\epsilon^2/n)$.

\subsection{Finding near-optimal strategies non-deterministically.}
Small simple strategies approximately determine the value of any game.
Thus, for exponentially large games, under the right conditions,
one can non-deterministically verify the approximate value.

\begin{theorem} \label{non-det strategy thm}
  Given $\epsilon>0$, $\rows$, $\cols$, 
  and an $\rows\times \cols$ game $\game$ 
  in the form of a poly-time oracle computing $\game_{\row\col}$ 
  from $\row$ and $\col$,
  the problem of computing for each player a mixed strategy
  that guarantees a payoff 
  within $\epsilon (\gameMax-\gameMin)$ of $\val{\game}$
  is in $\sigmaTwoOracle$,
  where $\poly(\gameOracle)$ means polynomial 
  in $\rows+\cols+\log(\rows\cols)/\epsilon$.
\end{theorem}
\begin{proof}
  We describe a $\sigmaTwoOracle$ computation
  that guesses and verifies the strategies for both players simultaneously,
  assuming WLOG that $\gameMax = 1$ and $\gameMin = 0$.

  Non-deterministically guess an approximate value $v$ for $\val{\game}$.
  By Theorem \ref{simple min-max},
  there is a strategy for $\Min$ that chooses uniformly
  from a multiset $S$ of $k$ pure strategies,
  where $k=O(\log (\cols)\,/\,\epsilon^2)$,
  and that guarantees a payoff less than $\val{\game}+\epsilon/2$.
  Guess $S$ non-deterministically.
  Use a non-deterministic oracle query to verify that
  \[
  (\forall \col) \sum_{\row\in S} \frac{1}{|S|} 
  \game_{\row\col} \le v+\epsilon/2.
  \]

  Similarly, guess and verify a mixed strategy for $\Max$
  that guarantees a payoff of at least $v-\epsilon/2$.

  Because each strategy guarantees an expected payoff
  within $\epsilon$ of that guaranteed by the other,
  each expected payoff is within $\epsilon$ of optimal.
\end{proof}

\subsection{Estimating circuit complexity 
  and generating hard distributions in $\sigmaTwo\cap\piTwo$.}

Using the close relationship established in Theorem \ref{anti-checker thm}
between circuit complexity and the value of the program/input game 
defined there, we can in some sense specialize the preceding theorem
to obtain the following result, which can be interpreted 
as showing that the circuit complexity of any NP-complete language
can be approximated within a linear factor in $\sigmaTwo\cap\piTwo$.
\begin{theorem} \label{verify thm}
  \begin{enumerate}
  \item 
    For any NP language $L$,
    there exists a decision procedure $A$ in $\sigmaTwo$
    such that $A(n,s)$ accepts if $\circComp_L(n) \ge 3ns$
    but rejects if $\circComp_L(n) \le s$.

  \item 
    For any NP-complete language L,
    there exists a decision procedure $B$ in $\sigmaTwo$
    such that $B(n,s)$ accepts if $\circComp_L(n) \le s$ 
    but rejects if $\circComp_L(n) > s$.
  \end{enumerate}

  Here the class $P$ is those languages
  decidable in time polynomial in $n$ and $s$.
\end{theorem}
\begin{proof}
  Let $L = \{x : (\exists y)\,\ell(x,y)\}$ 
  be defined by the poly($|x|$)-time predicate $\ell$.

  The decision procedure $A$ non-deterministically guesses 
  an anti-checker and uses standard techniques to verify it.
  By Theorem \ref{anti-checker thm},
  if $\circComp_L(n) \ge 3ns$, 
  then there are $O(s)$ inputs $\{x_i: i = 1,..,O(s)\}$
  such that any circuit of size $s$ misclassifies
  at least one input $x_i$.
  On the other hand, if $\circComp_L(n) \le s$
  then clearly no such set exists.
  Thus, the following predicate
  is true if $\circComp_L(n) \ge 3ns$
  but false if $\circComp_L(n) \le s$:
  \begin{eqnarray*}
    &&\left(\exists x_1,...,x_{O(s)}\right)\,
    \left(\forall C\right)\,\left(\exists i\right)\,
    \\&&~~(C(x_i)=0 \wedge (\exists y_i) \ell(x_i,y_i)=1)
    \\&&~~~\vee (C(x_i)=0 \wedge (\forall z_i)\,\ell(x_i,z_i)=0),
  \end{eqnarray*}
  where $C$ ranges over all size $s$ circuits and
  the $x_i$'s range over the $n$-bit strings.
  The ``$(\exists i)$'' quantifies over polynomially many $i$,
  so it can be expanded into an appropriate poly-size formula.  
  Using standard quantifier-elimination techniques,
  the resulting expression can be converted to the form
  $(\exists X,Y)(\forall C,Z) \ell'(n,s,C,X,Y,Z)$, 
  where $\ell'$ is a poly($n,s$)-time predicate.  
  Thus, the predicate is in $\sigmaTwo$.

  We construct the decision procedure $B(n,s)$ using standard techniques.
  It is known that, since $L$ is NP-complete, 
  the circuit complexity of the ``witness'' function $w$ 
  such that $\ell(x,w(x))$ for $x\in L$
  is only polynomially larger than $\circComp_L(n)$.
  Thus the following predicate holds
  iff $\circComp_L(n) \le s$.
  \begin{eqnarray*}
    && (\exists C,W) (\forall x)
    \\&&~~(C(x)=0 \wedge (\forall y) \ell(x,y)=0)
    \\&&~~~\vee (C(x)=1 \wedge \ell(x,W(x))=1),
  \end{eqnarray*}
  where $C$ ranges over all circuits of size $s$,
  $W$ ranges over all circuits
  of size large enough to compute the witness function,
  and $x$ ranges over all inputs of size $n$.
  This predicate is clearly in $\sigmaTwo$.
\end{proof}

\subsection{Hard distributions for uniform complexity classes.}
The first part of the proof of Theorem \ref{verify thm}
can easily be modified to show that
hard distributions for NP-complete languages
can be computed in $\sigmaTwo$.
This gives the following result.

\begin{proposition} \label{np prop}
Assume the polynomial-time hierarchy doesn't collapse to $\sigmaTwo$
and let $k>0$.
For any NP- or co-NP-complete language $L$ there is a distribution $D$ 
on $n$-bit strings such that
\begin{itemize}
  \item
    $D$ is computable in $\sigmaTwo$
  \item 
    no $O(n^k)$-time algorithm (even with $O(n^k)$ advice),
    when given a random input from $D$,
    decides membership in $L$ with probability of error less than $1/2-1/n^k$.
\end{itemize}
\end{proposition}
We leave the proof to the full paper.


\begin{thebibliography}{10}

\bibitem{Adleman:78}
Leonard~M. Adleman.
\newblock Two theorems on random polynomial time.
\newblock In {\em Proc. of the 19th IEEE Annual Symp. on Foundation of Computer
  Science}, pages 75--83, 1978.

\bibitem{Alon:Karp:Peleg:West:92}
Noga Alon, Richard~M. Karp, David Peleg, and Douglas West.
\newblock A graph-theoretic game and its application to the $k$-server problem.
\newblock In Lyle McGeoch and Daniel Sleator, editors, {\em On-Line Algorithms:
  Proceedings of a DIMACS Workshop}, volume~7 of {\em DIMACS Series in Discrete
  Mathematics and Theoretical Computer Science}, pages 1--9, 1992.

\bibitem{Althofer:94}
Ingo Alth{\"o}fer.
\newblock On sparse approximations to randomized strategies and covex
  combinations.
\newblock {\em Linear Algebra and its Applications}, 199, March 1994.

\bibitem{Ben-David:Chor:Goldreich:Luby:92}
Shai {Ben-David}, Benny Chor, Oded Goldreich, and Michael Luby.
\newblock On the theory of average case complexity.
\newblock {\em Journal of Computer and System Sciences}, 44:193--219, 1992.

\bibitem{Blum:Kannan:89}
Manuel Blum and S.~Kannan.
\newblock Designing programs that check their work.
\newblock In {\em Proc. of the 21st Ann. ACM Symp. on Theory of Computing},
  pages 86--97, 1989.

\bibitem{Chandra:Kozen:Stockmeyer:81}
Ashok~K. Chandra, Dexter Kozen, and Larry~J. Stockmeyer.
\newblock Alternation.
\newblock {\em Journal of the ACM}, 28(1):114--133, January 1981.

\bibitem{Feigenbaum:Koller:Shor:93}
Joan Feigenbaum, Daphne Koller, and Peter Shor.
\newblock Private communication.
\newblock 1993.

\bibitem{Goldwasser:Micali:Rackoff:89}
Shafi Goldwasser, Silvio Micali, and C.~Rackoff.
\newblock The knowledge complexity of interactive proof systems.
\newblock {\em SIAM Journal on Computing}, 18(1):186--208, 1989.

\bibitem{Gurevich:87}
Y.~Gurevich.
\newblock Complete and incomplete randomized {NP} problems.
\newblock In {\em Proc. of the 28th IEEE Annual Symp. on Foundation of Computer
  Science}, pages 111--117, 1987.

\bibitem{Gurevich:90}
Y.~Gurevich.
\newblock Matrix decomposition problem is complete for the average case.
\newblock In {\em Proc. of the 31st IEEE Annual Symp. on Foundation of Computer
  Science}, pages 802--811, 1990.

\bibitem{Hoeffding:63}
Wassily Hoeffding.
\newblock Probability inequalities for sums of bounded random variables.
\newblock {\em American Statistical Journal}, pages 13--30, March 1963.

\bibitem{Impagliazzo:Levin:90}
Russell Impagliazzo and Leonid Levin.
\newblock No better ways to generate hard {NP} instances than picking uniformly
  at random.
\newblock In {\em Proc. of the 31st IEEE Annual Symp. on Foundation of Computer
  Science}, pages 812--821, 1990.

\bibitem{Li:Vitanyi:89}
Ming Li and P.~M.~B. Vitanyi.
\newblock A theory of learning simple concepts under simple distributions and
  average case complexity for the universal distribution.
\newblock In {\em Proc. of the 30th IEEE Annual Symp. on Foundation of Computer
  Science}, pages 34--39, 1989.

\bibitem{Schapire:89}
Robert~E. Schapire.
\newblock The strength of weak learnability.
\newblock {\em Machine Learning}, 5:197--227, 1990.

\bibitem{Schoning:87}
Uwe Sch{\"{o}}ning.
\newblock Probabilistic complexity classes and lowness.
\newblock In {\em Proc. of the Second IEEE Structure in Complexity Theory
  Conference}, pages 2--8, 1987.

\bibitem{Venkatesan:Levin:88}
R.~Venkatesan and Leonid Levin.
\newblock Random instances of a graph coloring problem are hard.
\newblock In {\em Proc. of the 20th Ann. ACM Symp. on Theory of Computing},
  pages 217--222, 1988.

\bibitem{vonNeumann:28}
John von Neumann.
\newblock Zur {T}heorie der {G}esellschaftspiel.
\newblock {\em Mathematische Annalen}, 100(295-320), 1928.

\bibitem{Yao:77}
Andrew~C.C. Yao.
\newblock Probabilistic complexity: Towards a unified measure of complexity.
\newblock In {\em Proc. of the 18th IEEE Annual Symp. on Foundation of Computer
  Science}, pages 222--227, 1977.

\bibitem{Yao:83}
Andrew~C.C. Yao.
\newblock Lower bounds by probabilistic arguments.
\newblock In {\em Proc. of the 24th IEEE Annual Symp. on Foundation of Computer
  Science}, pages 420--428, 1983.

\bibitem{Young:94}
Neal~E. Young.
\newblock Greedy algorithms by derandomizing unknown distributions.
\newblock Technical Report T.R. 1087, Cornell University Department of
  Operations Research and Industrial Engineering, Ithaca, NY 14853, 1994.

\end{thebibliography}
\end{document}